\documentclass[two column,showpacs,preprintnumbers,amsmath,amssymb]{revtex4}
\usepackage{graphicx}
\usepackage{dcolumn}
\usepackage{bm}
\begin{document}

\title{Time-dependent barrier passage of Two-dimensional non-Ohmic damping system}

\author{Chun-Yang Wang}
\thanks{Corresponding author. Electronic mail: wchy@mail.bnu.edu.cn}
\affiliation{Department of Physics and engineering, Qufu Normal
University, Qufu, 273165, China}

%\date{\today}

\begin{abstract}
The time-dependent barrier passage of an anomalous damping system is
studied via the generalized Langevin equation (GLE) with non-Ohmic
memory damping friction tensor and corresponding thermal colored
noise tensor describing a particle passing over the saddle point of
a two-dimensional quadratic potential energy surface. The
time-dependent passing probability and transmission coefficient are
analytically obtained by using of the reactive flux method. The long
memory aspect of friction is revealed to originate a non-monotonic
$\delta$(power exponent of the friction) dependence of the passing
probability, the optimal incident angle of the particle and the
steady anomalous transmission coefficient. In the long time limit a
bigger steady transmission coefficient is obtained which means less
barrier recrossing than the one-dimensional case.
\end{abstract}

 \maketitle

\section{\label{sec:level1}INTRODUCTION}

The problem of escape from a metastable states potential is
ubiquitous in almost all scientific areas. A great amount of
chemical events, such as chemical reactions, molecular diffusion, or
collision of molecular systems, etc, can be modeled by a single
barrier escape process within the framework of standard Brownian
motion \cite{sbm1,sbm2,sbm3}. Although many other thoughts such as
transition state theory \cite{TST1,TST2,TST3} and unimolecular rate
theory \cite{urt1,urt2,urt3}, etc, still work, the celebrated
landmark elucidation of this problem is the Kramers rate theory
\cite{kramers}. Where in his famous work, H. A. Kramers established
a reaction rate formula which is applicable to all the cases from
moderate to strong damping. The method he used to calculate the rate
constant is conventional, namely, flux-over-population\cite{FOP}.
However, we noticed, among the various theoretical concepts for rate
calculation \cite{FOP,MFPT}, the method of reactive flux
\cite{JCP-Bao,rf1,rf2} is a most powerful and convenient way to
follow. In its spirit, the initial conditions are assumed to be at
the top of the barrier, which correspond to the ensemble of
trajectories which start with identical initial conditions but
experience different stochastic histories. The escape rate then can
be calculated by investigating the various flux of particles passing
the transition state at the saddle point. Achievements of its
application in one-dimensional (1D) cases \cite{JCP-Bao,rf1,rf2} has
been witnessed in the past decades, but it has never been used in
the study of higher dimensional systems.

 The recent widespread interest in anomalous
diffusion \cite{nsp1,nsp2,nsp3} has highlighted the critical role
that it plays in characterizing a large group of nonstandard
statistical physics. For which, the non-Ohmic model with a
idiosyncratic power law frequency depending spectral density
\cite{nonOhmic1,nonOhmic2,nonOhmic3} bears prominent consequences in
describing a rich variety of frequency-dependent damping mechanism.
Previous 1D studies on it have shown that in the non-Ohmic case, the
friction of the system has a long time memory, the transmission
coefficient of the system acts as a non-monotonic function of
$\delta$(power exponent of the friction) which reveals a relatively
strong recrossing phenomenon \cite{JCP-Bao,nsp3}. But no higher
dimensional study has occurred on this subject. However, a recent
work by us has shown that in high dimensional systems, the
non-diagonal correlation between various degrees of freedom will
endow the barrier passage a riveting character \cite{Chyw}, in
particular, particles moving in a two-dimensional (2D) potential
energy surface (PES) tend to select an optimal diffusion path to
surmount the barrier. So it is of great interest to investigate the
time-dependent barrier passage of high dimensional non-Ohmic damping
systems.

The primary purpose of this paper is then to report our recent study
on the non-Ohmic damping system. In Sec. II, the analytical
expression of the saddle-point passing probability is obtained by
solving the 2D coupled generalized Langevin equation with non-Ohimic
friction tensor. In Sec. III, we give the time-dependent
transmission coefficient derived by using of the reaction flux
method. Sec. IV serves as a summary of our conclusion.

\section{2D non-Ohmic passing probability}

We consider the directional diffusion of a particle in the 2D
quadratic PES: $U(x_1, x_2)=\frac{1}{2}\omega_{ij}x_ix_j$ with
$i,j=1,2$ and $\det\omega_{ij}<0$, the
 motion of the particle is described by the generalized Langevin equation (GLE) with
a non-Ohmic memory friction tensor:
\begin{equation}
m_{ij}\ddot{x}_{j}(t)+\int_{{0}}^{\infty}dt'\beta_{ij}(t-t')\dot{x}_{j}(t')+\omega_{ij}x_{j}(t)=\xi_{i}(t),\label{eq,GLE}
\end{equation}
where the Einstein summation convention is used and the components
of the random force are assumed to be zero-mean and their
correlations obey the fluctuation-dissipation theorem
\begin{eqnarray}
\langle\xi_{i}(t)\xi_{j}(t')\rangle&=&k_{_{B}}T\beta_{ij}(t-t')
\nonumber\\&=&\frac{2k_{_{B}}T}{\pi}\int_{0}^{\infty}d\omega\frac{J_{ij}(\omega)}{\omega}\cos\omega(t-t'),
\end{eqnarray}%
with $k_{_{B}}$ the Boltzmann constant, $T$ the temperature of the
reservoir and $J_{ij}(\omega)$ takes the form of non-Ohmic spectral
density $J_{ij}(\omega)=\gamma_{ij}(\omega/\omega_{r})^{\delta}$,
where $\delta$ is the power exponent taking values between 0 and 2,
$\gamma_{ij}$ is the friction constant, and $\omega_{r}$ denotes a
reference frequency allowing for the damping constant $\gamma_{ij}$
to have the dimension of a viscosity at any $\delta$.

The initial condition of the particle is denoted as
$x_{j}(0)=x_{j0}$ and $\dot{x}_j(0)=v_{j0}$, where $x_{10}<0$ and
$v_{10}>0$. Assuming that $x_{1}$-axis is the transport direction
[$\omega_{11}<0$], the reduced distribution function of the particle
for $x_1$, when the variables $x_{2}(t)$, $v_{1}(t)$ and $v_{2}(t)$
are integrated out, can be written as:
\begin{eqnarray}
&&W(x_{1},t; x_{_{10}},x_{_{20}},v_{_{10}},v_{_{20}})\nonumber\\
&& \ \ \ \ \ \ \ \ \
=\frac{1}{\sqrt{2\pi}\sigma_{x_{1}}(t)}\textrm{exp}\left({-\frac{(x_{1}-\langle
x_{1}(t)\rangle)^{2}}{2\sigma^{2}_{x_{1}}(t)}}\right).\label{eq,pd}
\end{eqnarray}
Applying on it the Laplace transform technique to Eq.(\ref{eq,GLE})
we get $x_{1}(t)$ and its variance $\sigma^{2}_{x_{1}}(t)$ at any
time
\begin{eqnarray}
x_{1}(t)&=&\langle
x_{1}(t)\rangle+\sum^{2}_{i=1}\int^{t}_{0}H_{i}(t-t')\xi_{i}(t')dt',\label{eq,p1}\\
\sigma^{2}_{x_{1}}(t)&=&\int^{t}_{0}dt_{1}H_{i}(t-t_{1})\int^{t_{1}}_{0}dt_{2}\langle
\xi_{i}(t_{1})\xi_{j}(t_{2})\rangle
H_{j}(t-t_{2}),\label{eq,p2}\nonumber\\
\end{eqnarray}
where the mean position of the particle along the transport
direction is given by
\begin{equation}
\langle
x_{1}(t)\rangle=\sum^{2}_{i=1}\left[C_{i}(t)x_{i0}+C_{i+2}(t)v_{i0}\right],\label{eq,ip}
\end{equation}
which relates to the initial position and velocity. The
time-dependent factors in Eq.(\ref{eq,ip}) according to the residual
theorem are
$C_{i}(t)=\mathcal{L}^{-1}[\textsl{F}_{i}(s)/\textsl{P}(s)]$
$(i=1\ldots4)$ with
 exponential forms, the two response functions in Eqs.(\ref{eq,p1}) and (\ref{eq,p2}) read
$\textsl{H}_{1}(t)=\mathcal{L}^{-1}[\textsl{F}_{5}(s)/\textsl{P}(s)]$
and
$\textsl{H}_{2}(t)=\mathcal{L}^{-1}[\textsl{F}_{6}(s)/\textsl{P}(s)]$,
where $\mathcal{L}^{-1}$ denotes the inverse Laplace transform. The
expressions of $P(s)$ and $F_{i}(s)$ $(i=1,...,6)$  are written in
the appendix.

The probability of passing over the saddle point [$x_1=x_2=0$] is
always called the characteristic function
$\chi(x_{_{10}},x_{_{20}},v_{_{10}},v_{_{20}};t)$ which can be
determined mathematically by integrating Eq.(\ref{eq,pd}) over
$x_{1}$ from zero to infinity
\begin{eqnarray}
\chi(
x_{_{10}},x_{_{20}},v_{_{10}},v_{_{20}};t)&=&\int^{\infty}_{0}W(x_{1},t;
x_{_{10}},x_{_{20}},v_{_{10}},v_{_{20}})dx_{1}\nonumber\\
&=&\frac{1}{2}\textrm{erfc}\left(-\frac{\langle
x_{1}(t)\rangle}{\sqrt{2}\sigma_{x_{1}}(t)}\right).\label{eq,chsi}
\end{eqnarray}
In the barrier surmounting process, the probability density function
of the particle that is originally at one side of the barrier
spreads in the phase space due to thermal fluctuation, and its
center moves driven by its initial velocity. For large times, the
probability of passing over the saddle point converges to a finite
value depending on the initial position and velocity of the
particle, and the other fraction of the particles goes to infinite
along the opposite direction. This results in the characteristic
function Eq.(\ref{eq,chsi}) to be a number which is equal to 1 for
reactive trajectories and 0 for nonreactive trajectories. If we
consider only the long time result of state transition induced by
the particle diffusion, the stationary passing probability is of
great important to rely on. Mathematically it is calculated by
$P_{\textmd{pass}}=\lim_{t\to\infty}\frac{1}{2}\textmd{erfc}\left[-\langle
x_{1}(t)\rangle/(\sqrt{2}\sigma_{x_1}(t))\right]$.

\begin{figure}
%\centering
\includegraphics[scale=0.75]{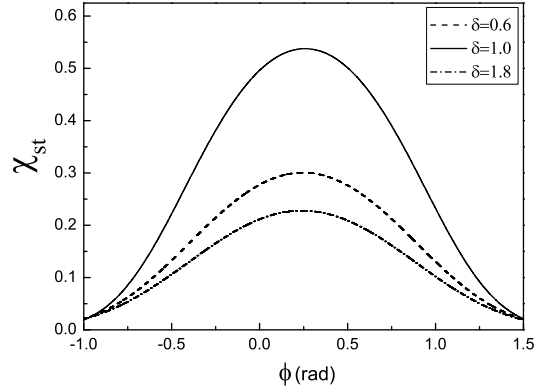}
\caption{Stationary passing probability as a function of the
incident angel for various $\delta$. The parameters used are
$\omega_{12}=-0.5$, $m_{12}=0.6$, $\omega_{11}=-2.0$,
$\omega_{22}=1.5$, $m_{11}=1.5$, $m_{22}=2.0$, $\gamma_{12}=1.0$,
$\gamma_{11}=1.8$, $\gamma_{22}=1.2$, $\omega_{r}=1.0$, and the
initial condition of the particle is set $x_{10}=-1.0$, $x_{20}=0$,
$v_{0}=1.9$.\label{Fig1}}
\end{figure}

In Fig. \ref{Fig1}, we plot the stationary passing probability as a
function of the incident angel for various power exponents $\delta$.
In the calculations here and following, we rescale all the variables
so that the dimensionless unit such as $k_{B}=1.0$ is used. Each
friction strength $\gamma_{ij}$ is fixed to be a constant
independent of $\delta$, and the reference frequency
$\omega_{r}=1.0$. It is seen from Fig. \ref{Fig1} that, the
stationary passing probability of non-Ohmic case ($\delta=0.6$ or
$\delta=1.8$) is not so large as the result of Ohmic case
($\delta=1.0$). This can be understood from the point of view of the
particle's critical velocity which corresponds to the condition:
$\lim_{t\rightarrow\infty}\langle x_{1}(t)\rangle=0$. Supposing
$v_{_{10}}=v_{_{0}}\textrm{cos}\phi$ and
$v_{_{20}}=v_{_{0}}\textrm{sin}\phi$ are the two components of the
particle's velocity along two directions $x_{1}$ and $x_{2}$. The
critical velocity for a particle reads:
\begin{eqnarray}
v^{c}_{_{0}}=-\frac{C_{1}(\infty)x_{_{10}}+C_{2}(\infty)x_{_{20}}}
{C_{3}(\infty)\textrm{cos}\phi+C_{4}(\infty)\textrm{sin}\phi}.\label{eq,cv}
\end{eqnarray}
For the parameters in Fig.\ref{Fig1}, we obtain
$v^{c}_{_{0}}|_{\delta=1.8}\cong2.6417>v^{c}_{_{0}}
|_{\delta=0.6}\cong2.3477>v^{c}_{_{0}}|_{\delta=1.0}\cong1.8426$
when the incident angle is set $\phi=0.258 \textmd{rad}$. Thus we
can see the particle of non-Ohmic damping system has a small passing
probability.

It is also shown in Fig. \ref{Fig1} that, for each $\delta$ there
exits a certain incident angle for the particle to reach a maximum
stationary passing probability. This implies, for the 2D non-Ohmic
damping system, an optimal incident angle (or an optimal diffusion
path) still exists for the particle to surmount the PES, in
accordance with the Ohmic case ($\delta=1.0$) we have studied
\cite{Chyw}. The optimal incident angle for a particle of 2D
non-Ohmic damping system can also be determined by retrospect the
minimum value of the critical velocity Eq. (\ref{eq,cv}). After some
algebra we find in the long time limit it can be expressed by the
system parameters as
\begin{eqnarray}
\psi=\textrm{arctan}\left(\frac{m_{12}(\hat{\beta}_{22}[s]\varepsilon+\omega_{22})-m_{22}(\hat{\beta}_{12}[s]\varepsilon+\omega_{12})
}{m_{11}\textsl{F}_{5}(\varepsilon)+m_{12}\textsl{F}_{6}(\varepsilon)}\right).\label{eq,oa}
\end{eqnarray}
where $\varepsilon$ is the largest analytical root of $P(s)=0$,
$\hat{\beta}_{ij}[s]=\tilde{\gamma}_{ij}s^{\delta-1}$ is the Laplace
transform of the friction kernel function and
$\tilde{\gamma}_{ij}=\gamma_{ij}\omega_{r}^{1-\delta}\textrm{sin}^{-1}(\delta\pi/2)$
is namely the effective friction here.

\begin{figure}
%\centering
\includegraphics[scale=0.75]{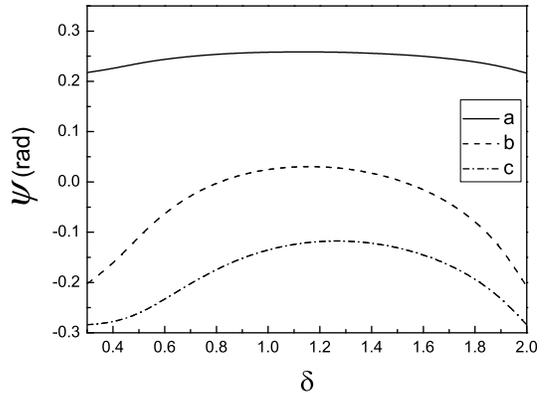}
\caption{Optimal incident angel as a function of the power exponents
$\delta$ for various effective frictions. The different parameters
for each curve are (a) $\gamma_{12}=0.8$, $\gamma_{11}=1.8$,
$\gamma_{22}=1.2$; (b) $\gamma_{12}=2.0$, $\gamma_{11}=2.5$,
$\gamma_{22}=2.2$; (c) $\gamma_{12}=4.0$, $\gamma_{11}=4.8$,
$\gamma_{22}=4.2$, others are the same as those used in Fig. 1.
\label{Fig2}}
\end{figure}

Noticing that Eq. (\ref{eq,oa}) implicitly implies the power
exponent $\delta$. This means in the non-Ohmic damping case the
optimal incident angle has an intimate relation to the friction of
the system. In Fig. \ref{Fig2}, the $\delta$ dependence of the
optimal incident angle $\psi$ at various effective friction is
plotted. It shows that $\psi$ evolves as a non-monotonic function of
$\delta$. This implies that the barrier passage of the
two-dimensional non-Ohmic damping system is intimately controlled by
the friction. Particle facing different friction strength will
select different optimal path in its barrier surmounting process.

\begin{figure}
%\centering
\includegraphics[scale=0.9]{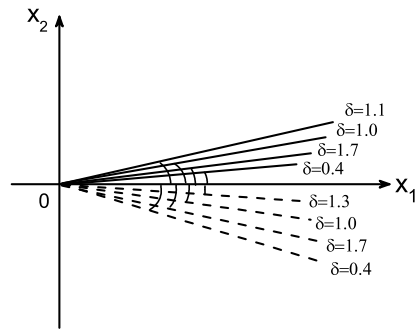}
\caption{Schematic illustration of the optimal incident angel for
various power exponents $\delta$. The different parameters for each
group of curves are (a) $\gamma_{12}=0.8$, $\gamma_{11}=1.8$,
$\gamma_{22}=1.2$ for straight lines; and (b) $\gamma_{12}=4.0$,
$\gamma_{11}=4.8$, $\gamma_{22}=4.2$ for dashed lines, other
parameters are the same as those used in Fig. 1.\label{figh}}
\end{figure}

For instance Fig. \ref{figh} gives a schematic illustration of the
optimal incident angel for various power exponents $\delta$ at
different friction strength. From which we can see, comparing with
the Ohmic case (i.e. $\delta=1.0$), when the system friction
strength is not so strong (straight lines in Fig. \ref{figh}), the
optimal incident angle of non-Ohmic case tends to approach zero from
a clockwise direction at most values of $\delta$. However, if the
system has a relatively strong friction (dashed lines in Fig.
\ref{figh}), the optimal incident angle of non-Ohmic case tends to
approach zero from a anticlockwise direction. Since the direction of
zero incident angle represents the valley direction of the PES, this
implies the particle diffusion under the influence of non-Ohmic
friction will has the probability of tracing along the valley
passage. This is a meaningful prediction for the study of barrier
escape problem. To deicide the relation between the system
parameters that enables the particle to travel along the valley
direction, one needs only set $\psi=0$. This will be very useful in
studying the fusion reactions of massive nuclei.

\section{non-Ohmic Anomalous transmission coefficient}

In this section we evaluate the rate of a particle escape from the
2D metastable potential. Since difficulty makes it impossible to
solve the Fokker-Planck equation with non-Ohmic friction, we use the
method of reactive flux to derive the rate constant from Langevin
dynamics. In the spirit of reactive flux calculation, particles are
set evolving from the top of the barrier and the thermal rate
constant is got from the ensemble average of trajectories starting
with identical initial conditions but experience different
stochastic histories. For a 2D system, the rate for a particle of
unit mass \cite{2Drate} is defined as
\begin{eqnarray}
k(t)=\frac{1}{Qh}\int d\vec{p}ds(\vec{p}\cdot
\vec{n}_{s})\chi(\vec{p},\vec{q}_{s})e^{-H(\vec{p},\vec{q}_{s})/k_{_{B}}T}.
\end{eqnarray}
Where $\vec{p}$ is the momentum vector and $Q$ is the partition
function for reactants integrating over the distribution of the
ground states. $\chi(\vec{p},\vec{q}_{s})$, namely the
characteristic function is a number which is equal to 1 for reactive
trajectories ($\vec{p} \cdot \vec{n}_{s}>0$) and 0 for nonreactive
ones ($\vec{p} \cdot \vec{n}_{s}<0$). Because of the stochastic
nature of the dynamics, it is necessary to take into account all the
different possible realizations of a trajectory. To do so
$\chi(\vec{p},\vec{q}_{s})$ should be replaced by its
non-equilibrium average
$\chi(x_{_{10}},x_{_{20}},v_{_{10}},v_{_{20}})$ which results in a
weighted factor with its value ranging from 0 to 1. The rate is then
proportional to the total flux from reactants to products
independent of the choice of dividing surface between reactants and
products excluding the influence of the recrossing effects.

If the initial conditions are assumed to be at the top of the
barrier, the rate constant corresponds to an ensemble of particles
starting from ($x_{_{10}}=0,x_{_{20}},v_{_{10}},v_{_{20}}$) at $t=0$
and obeys the equilibrium distribution. Mathematically,
\begin{widetext}
\begin{eqnarray}
k(t)&=&\frac{1}{Qh}\prod^{2}_{i=1}\int^{\infty}_{-\infty}dx_{_{i0}}
dv_{_{i0}}v_{_{10}}
e^{-H(x_{_{10}},x_{_{20}},v_{_{10}},v_{_{20}})/k_{_{B}}T}
\delta(x_{_{10}})\chi(x_{_{10}},x_{_{20}},v_{_{10}},v_{_{20}};t)\nonumber\\
&=&\frac{e^{-V_{_{B}}/k_{B}T}}{Qh}\prod^{2}_{i=1}\int^{\infty}_{-\infty}dx_{_{i0}}
dv_{_{i0}}v_{_{10}}
e^{-\left[m_{ij}v_{_{i0}}v_{_{j0}}+\omega_{22}x^{2}_{_{20}}\right]/2k_{_{B}}T}
\chi(x_{_{10}}=0,x_{_{20}},v_{_{10}},v_{_{20}};t),\label{eq,rate}
\end{eqnarray}
with the system Hamiltonian
$H(x_{_{10}},x_{_{20}},v_{_{10}},v_{_{20}})
=\frac{1}{2}m_{ij}v_{_{i0}}v_{_{j0}}+\frac{1}{2}\omega_{ij}x_{_{i0}}x_{_{j0}}+V_{_{B}}$,
and an initial Boltzmann form stationary probability distribution
with a weight $k_{B}T$ at the metastable well is assumed when the
temperature is much less than the barrier height of the metastable
potential.

 The total rate thus can be viewed as a transition state theory
(TST) rate $k^{TST}=\frac{1}{Qh}e^{-V_{_{B}}/k_{B}T}$
\cite{TST1,TST2,TST3} multiplied by a factor $\kappa(t)$ between 0
and 1, namely the transmission coefficient, as we have got here
\begin{eqnarray}
\kappa(t)=\left[1+\frac{\textrm{det}m_{ij}}{m_{22}}\left(\frac{C^{2}_{2}(t)}{\omega_{22}}-\frac{2m_{12}}
{\textrm{det}m_{ij}}C_{3}(t)C_{4}(t)+\frac{m_{11}C^{2}_{4}(t)}{\textrm{det}m_{ij}}+\frac{\sigma^{2}_{x_{1}}(t)}{k_{B}T}\right)\right]^{-\frac{1}{2}}.
\label{eq,tansco}
\end{eqnarray}
\end{widetext}
which describes the probability of a particle successfully escaped
from the metastable well to recross the barrier. This expression for
the fractional reactive index leads immediately to Kramers¡¯ formula
for the rate constant.

\begin{figure}
%\centering
\includegraphics[scale=0.75]{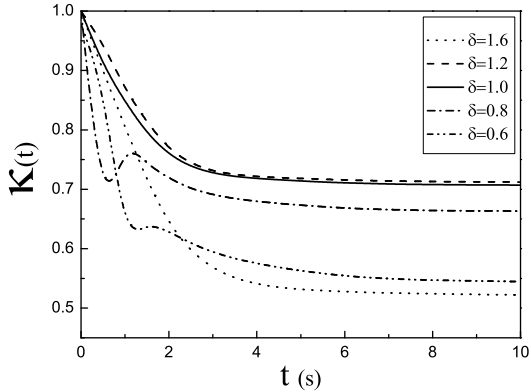}
\caption{Transient expression of $\kappa(t)$ for various strength of
system friction. Identical parameters are used as in Fig.
\ref{Fig1}.}\label{Fig4}
\end{figure}

In order to individually inspect the dynamical corrections of
$\kappa(t)$ to the TST rate, Fig. \ref{Fig4} gives the transient
expression of it for various strength of system friction. Apparently
seen from it, the value of $\kappa(t)$ keeps $1.0$ when $t=0$,
showing no barrier recrossing behavior at the beginning. But as
$t\rightarrow\infty$, $\kappa(t)$ stabilizes to a invariable
constant between 0 and 1, namely the stationary transmission
coefficient $\kappa_{\textrm{st}}$, representing the probability of
a particle already escaped from the metastable potential well to
recross the barrier. Mathematically, it creates a cutoff factor to
the TST rate.

Meanwhile, by investigating the stationary transmission coefficient
at various strength of system friction we can find: in most cases of
non-Ohmic damping, $\kappa(t)$ stabilizes to a smaller
$\kappa_{\textrm{st}}$ than the Ohmic case ($\delta=1.0$) except for
some special one such as $\delta=1.2$. This implies the barrier
recrossing of non-Ohmic system is generally stronger than the Ohmic
case. The long memory aspect of friction is most possible to be at
the origin of this behavior.

\begin{figure}
%\centering
\includegraphics[scale=0.75]{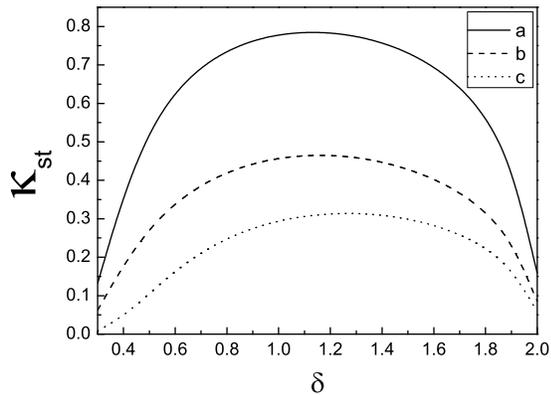}
\caption{Stationary transmission coefficient $\kappa_{\textrm{st}}$
as a function of $\delta$ for various system frictions. Parameters
used are identical to those in Fig \ref{Fig2}.\label{Fig5}}
\end{figure}

For further understanding of this subject, the stationary
transmission coefficient $\kappa_{\textrm{st}}$ is plotted in Fig.
\ref{Fig5} as a function of $\delta$ for various system frictions.
Again it is found to change non-monotonically with $\delta$. This is
because, the friction of non-Ohmic system has a long memory in time.
If the friction strength is weak (i.e. $\delta\rightarrow0$), the
motion of a particle depending greatly on the initial conditions. In
the opposite aspect, strong friction strength (i.e.
$\delta\rightarrow2.0$) causes great viscous force to the particle.
In both cases, it is unfavorable for the particle to pass the saddle
point of the PES. This results in a smaller passing probability and
a transmission coefficient than the normally Ohmic case.

Furthermore, comparing with the results of 1D non-Ohmic
systems\cite{JCP-Bao,jlz}, the 2D non-Ohmic system is found to have
a bigger stationary transmission coefficient in the long time limit
which means a weaker barrier recrossing effect. Particle diffusing
in the 2D PES under non-Ohmic friction seldom returns after passing
the saddle point of the potential. Macroscopically it results in a
big net flux rate.

\section{SUMMARY}

In conclusion, a two-dimensional barrier passage has been
investigated, where the particle is subjected to a non-Ohmic memory
friction with power exponent $\delta$. The time-dependent passing
probability and transmission coefficient are analytically obtained
by using of the reaction flux method in a Langevin dynamic picture.
In the long time limit, the passing probability, the optimal
incident angle of the particle and the anomalous transmission
coefficient are all found to be non-monotonic functions of the power
exponent of the friction $\delta$. The long memory aspect of
friction is most possible to be at the origin of this behavior. The
relatively bigger stationary transmission coefficient reveals that
there is less barrier recrossing in the two-dimensional system than
the one-dimensional case. It is very difficult for the particle
diffusing in the 2D PES under non-Ohmic friction to return after
passing the saddle point of the potential. Thus results in a big net
flux rate.

\section * {ACKNOWLEDGEMENTS}
This work was supported by the Scientific Research Starting
Foundation of Qufu Normal University and the National Natural
Science Foundation of China under Grant No. 10847101.

\section * {APPENDIX. THE EXPRESSIONS OF $\langle x_1(t)\rangle$}

The quantities appear in the expression of $\langle x_1(t)\rangle$
are
\begin{eqnarray}
\textsl{P}(s)&=&(m_{11}\hat{\beta}_{22}[s]+m_{22}\hat{\beta}_{11}[s]-2m_{12}\hat{\beta}_{12}[s])s^{3}
\nonumber\\&&
+(\textrm{det}\hat{\beta}_{ij}[s]+m_{11}\omega_{22}+m_{22}\omega_{11}-2m_{12}\omega_{12})s^{2}\nonumber\\&&
+(\hat{\beta}_{11}[s]\omega_{22}+\hat{\beta}_{22}[s]\omega_{11}-2\hat{\beta}_{12}[s]\omega_{12})s\nonumber\\&&+(\textrm{det}m_{ij})s^{4}
+\textrm{det}\omega_{ij},\nonumber\\
\textsl{F}_{1}(s)&=&(m_{11}\hat{\beta}_{22}[s]+m_{22}\hat{\beta}_{11}[s]-2m_{12}\hat{\beta}_{12}[s])s^{2}
\nonumber\\&&
+(\textrm{det}\hat{\beta}_{ij}[s]+m_{11}\omega_{22}-m_{12}\omega_{12})s+(\textrm{det}m_{ij})s^{3}\nonumber\\&&
+\hat{\beta}_{11}[s]\omega_{22}-\hat{\beta}_{12}[s]\omega_{12},\nonumber\\
\textsl{F}_{2}(s)&=&(m_{12}\omega_{22}-m_{22}\omega_{12})s+\hat{\beta}_{12}[s]\omega_{22}-\hat{\beta}_{22}[s]\omega_{12},\nonumber\\
\textsl{F}_{3}(s)&=&(\textrm{det}m_{ij})s^{2}+(m_{11}\hat{\beta}_{22}[s]-m_{12}\hat{\beta}_{12}[s])s
\nonumber\\
&&+m_{11}\omega_{22}-m_{12}\omega_{12},\nonumber\\
\textsl{F}_{4}(s)&=&(m_{12}\hat{\beta}_{22}[s]-m_{22}\hat{\beta}_{12}[s])s
+m_{12}\omega_{22}-m_{22}\omega_{12},\nonumber\\
\textsl{F}_{5}(s)&=&m_{22}s^{2} +\hat{\beta}_{22}[s]s+\omega_{22},\nonumber\\
\textsl{F}_{6}(s)&=&-m_{12}s^{2}-\hat{\beta}_{12}[s]s-\omega_{12}.\nonumber
\end{eqnarray}
where $\hat{\beta}_{ij}[s]=\tilde{\gamma}_{ij}s^{\delta-1}$ is the
Laplace transform of the friction kernel function and
$\tilde{\gamma}_{ij}=\gamma_{ij}\omega_{r}^{1-\delta}\textrm{sin}^{-1}(\delta\pi/2)$
is namely the effective friction.


\begin{thebibliography}{99}

\bibitem{sbm1}P. H\"{a}nggi, P. Talkner, and M. Borkovec, Rev. Mod. Phys. \textbf{62}, 251 (1990).

\bibitem{sbm2} E. Pollak, J. Chem. Phys. \textbf{85}, 865 (1986).

\bibitem{sbm3} P. Talker, E. Pollak, and A. M. Berhkovskii, Chem. Phys. \textbf{235}, 1 (1998).

\bibitem{TST1} T. Seideman and W. H. Miller, J. Chem. Phys. \textbf{95}, 1768 (1991).

\bibitem{TST2} J. M. Sancho, A. H. Romero, and K. Lindenberg, J. Chem. Phys. \textbf{109}, 9888 (1998).

\bibitem{TST3} E. Pollak and M. S. Child, J. Chem. Phys. \textbf{72}, 1669 (1980).

\bibitem{urt1} W. Forst, \textit{Theory of Unimolecular Reaction} (Academic, New York, 1973).

\bibitem{urt2} W. L. Hase, in \textit{Dynamics of Molecular Collisions}, Part B, edited by W. H.
Miller (Plenum, New York, 1976), p. 121.

\bibitem{urt3} A. B. Callear, in \textit{Modern Methods
in Kinetics}, Comprehensive Chemical Kinetics, Vol. 24, edited by C.
H. Bamford and C. F. H. Tipper (Elsevier, New York, 1983) p.333.

\bibitem{kramers} H. A. Kramers, Physica (Utrecht) \textbf{7}, 284 (1940).

\bibitem{FOP} L. Farkas, Z. Phys. Chem. \textbf{125}, 236 (1927).

\bibitem{MFPT} P. Talkner, Z. Phys. B \textbf{68}, 201 (1987).

\bibitem{JCP-Bao} J. D. Bao, J. Chem. Phys. \textbf{124}, 114103 (2006)

\bibitem{rf1} D. J. Tannor and D. Kohen, J. Chem. Phys. \textbf{100}, 4932 (1994).

\bibitem{rf2} D. Kohen and D. J. Tannor, J. Chem. Phys. \textbf{103}, 6013 (1995).

\bibitem{nsp1} R. Metzler and J. Klafter, Phys. Rep. \textbf{339}, 1 (2000).

\bibitem{nsp2} J. D. Bao and Y. Z. Zhuo, Phys. Rev. Lett.\textbf{ 91}, 138104 (2003).

\bibitem{nsp3} Jing-Dong Bao, Yi-Zhong Zhuo. Phys. Rev. C \textbf{67}, 064606 (2003).

\bibitem{nonOhmic1} U. Weiss, \textit{Quantum Dissipative Systems}, 2nd ed. (World Scientific, Singapore,
1999).

\bibitem{nonOhmic2} N. Pottier, Physica A \textbf{317}, 371 (2003).

\bibitem{nonOhmic3} H. Grabert, P. Schramm, and G.-L. Ingold, Phys. Rev. Lett. \textbf{58}, 1285 (1987).



\bibitem{Chyw} C. Y. Wang, Jia Y and J. D. Bao, Phys. Rev. C \textbf{77}, 024603 (2008)

\bibitem{2Drate} P. Pechukas, in \textit{Dynamics of Molecular Collisions}, Part B, edited by W. H.
Miller (Plenum, New York, 1976), p. 269.

\bibitem{jlz} J. L. Zhao and J. D. Bao, Phys. A \textbf{356}, 517 (2005)


\end{thebibliography}
\end{document}